\documentclass[a4paper,11pt]{article}
\pdfoutput=1 

\usepackage{jheppub} 
\usepackage{lineno}
\usepackage[T1]{fontenc} 
\usepackage{appendix}
\usepackage{amsmath}
\usepackage{slashed}
\usepackage{subcaption}
\usepackage{orcidlink}

\newcommand{\Lagrangian[1]}{\mathcal{L}_{#1}}

\title{\boldmath Discriminating Majorana and Dirac heavy neutrinos at lepton colliders}

\author[a,b,
\orcidlink{0000-0003-4268-508X}]{Krzysztof~M\k{e}ka{\l}a,}
\author[b,
\orcidlink{0000-0003-1866-0157}]{J\"urgen Reuter\;}
\author[a,
\orcidlink{0000-0001-8975-9483}]{and Aleksander Filip~\.Zarnecki\;}
\affiliation[a]{Faculty of Physics, University of Warsaw,
                Pasteura 5, 02-093 Warszawa, Poland}
\affiliation[b]{Deutsches Elektronen-Synchrotron DESY, Notkestr. 85, 22607 Hamburg, Germany}

\emailAdd{k.mekala@uw.edu.pl}
\emailAdd{juergen.reuter@desy.de}
\emailAdd{zarnecki@fuw.edu.pl}

\abstract{In this paper we investigate how well the nature of heavy neutral leptons can be determined at a future lepton collider, after its potential 
discovery. Considered in a simplified model are prompt decays of the neutrino in the mass range from 100 GeV to 10 TeV. We study event selection and application of multivariate analyses to determine whether such a newly discovered particle is of the Dirac or Majorana nature. Combining lepton charge and kinematic event variables,
we find that the nature of a heavy neutrino, whether it is a Dirac or a 
Majorana particle, can be determined at 95\% C.L. almost in the whole discovery range. We will briefly speculate about other than the studied channels and the robustness of this statement in
more general models of heavy neutral leptons, particularly on the complementarity of high-energy electron-positron vs. muon colliders on resolving the flavor structure of heavy neutrinos.}

\begin{document}
\maketitle
\flushbottom

\section{Introduction}
\label{sec:intro}

The discovery of neutrino masses in neutrino oscillation was the first firm sign of physics beyond the Standard Model (SM), connected to lepton flavor violation (LFV) in the neutral lepton sector~\cite{Gonzalez-Garcia:2002bkq,Gonzalez-Garcia:2007dlo}. Neutrino masses are nowadays experimentally known to be five orders below the electron mass and eleven orders of magnitude below the top-quark mass~\cite{ParticleDataGroup:2022pth,Fogli:2005cq}. These tiny masses together with the phenomenon of quasi-maximal lepton-flavor mixing angles have given neutrinos the status of the most mysterious particles known to date. A potential explanation for these tiny masses, the seesaw mechanism~\cite{Minkowski:1977sc,Mohapatra:1980yp,Yanagida:1980xy,Schechter:1981cv}, has been established. Sectors of heavy Majorana or Dirac neutrinos could mix with the degrees of freedom with electroweak (EW) quantum numbers, resembling Standard Model neutrinos. These sectors could provide additional CP violation necessary to explain the matter-antimatter asymmetry of the universe, candidates for dark matter or potentially first-order phase transitions within the early universe. The most minimal models point either to very high-scale neutrinos masses of the order of $10^{14}$ GeV or above, or admit specific symmetries leading, for example, to neutrinos appearing in signatures of displaced vertices~\cite{ATLAS:2020xyo,CMS:2022fut,Drewes:2019fou} or so-called pseudo-Dirac neutrinos (mixtures of Dirac and Majorana neutrinos)~\cite{Antusch:2022ceb,Antusch:2022hhh}. However, without experimental confirmation of any of these UV-complete models, it is justified to construct effective field theories or simplified models that exhibit the minimal set of features of UV-complete models to test them at present or future colliders\footnote{One proposal of a TeV-scale neutrino has been made e.g. in the context of Little Higgs models~\cite{Kilian:2003xt}.}. Once heavy neutrinos (or \textit{heavy neutral leptons}, HNLs) have been discovered, we need to determine their properties to reconstruct the UV-complete model: their masses, their widths, their mixing parameters with the light neutrinos, their spin and whether they carry a complex or real representation of the Lorentz group, i.e. they are Dirac or Majorana spinors. For this purpose, we take a simplified model setup with heavy neutrinos (either Dirac or Majorana) that mix with SM neutrinos and provide a framework to reveal their nature at future lepton (electron-positron and muon) colliders.

\begin{figure}[tb]
    \centering
    \includegraphics[width=0.6\textwidth]{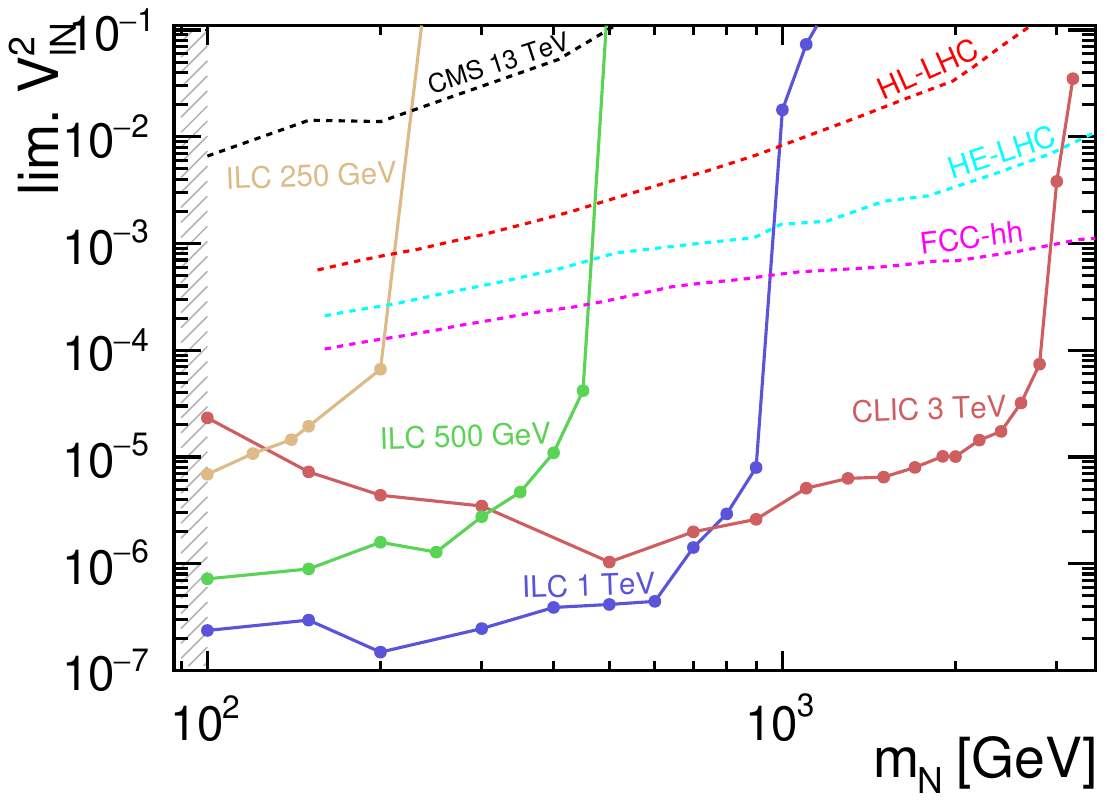}
    \caption{Exclusion limits on the coupling $V^2_{l N}$, as a function of the heavy Dirac neutrino mass, $m_N$, for different lepton collider setups (solid lines), as indicated in the plot.
    Dashed lines indicate limits from current and future hadron colliders based on \cite{Sirunyan:2018mtv,Pascoli:2018heg}. The plot is taken from \cite{Mekala:2022cmm} and updated to include results for ILC250~\cite{talk_ECFA}.
    \label{fig:results}
    }
\end{figure}

The signatures of the heavy neutrinos at lepton colliders have already been broadly discussed in the literature~\cite{delAguila:2005pin, delAguila:2005ssc, Saito:2010xj, Das:2012ze, Antusch:2016ejd, Banerjee:2015gca, Chakraborty:2018khw, Das:2018usr, Cai:2017mow, Mekala:2022cmm, Mekala:2023diu, Kwok:2023dck, Li:2023tbx, Antonov:2023otp, Mikulenko:2023ezx}. In \cite{Mekala:2022cmm, Mekala:2023diu}, we presented an efficient way to search for prompt decays of HNLs at future machines and to set limits on their mixing parameter assuming that they are the only BSM particles and they couple equally to all the SM leptons. 
Figure \ref{fig:results}, taken from \cite{Mekala:2022cmm} and updated to account for new results for ILC running at 250 GeV\footnote{We actually expect very similar results for other Higgs factories at the $ZH$ stage, e.g. FCC-ee or CEPC at 240 GeV, but we did not perform a dedicated study based on the IDEA or CLD detectors.}, presents exclusion limits on the mixing parameter which could be set by future $e^+e^-$ machines. 
Comparison of the expected lepton collider limits with results for $pp$ machines shows that the former would perform much better in this regard, superseding FCC-hh running at 100\,TeV by orders of magnitude for masses up to the collision energy.

\begin{figure}[tb]
    \centering
    \begin{subfigure}{0.49\textwidth}
    \includegraphics[width=\textwidth]{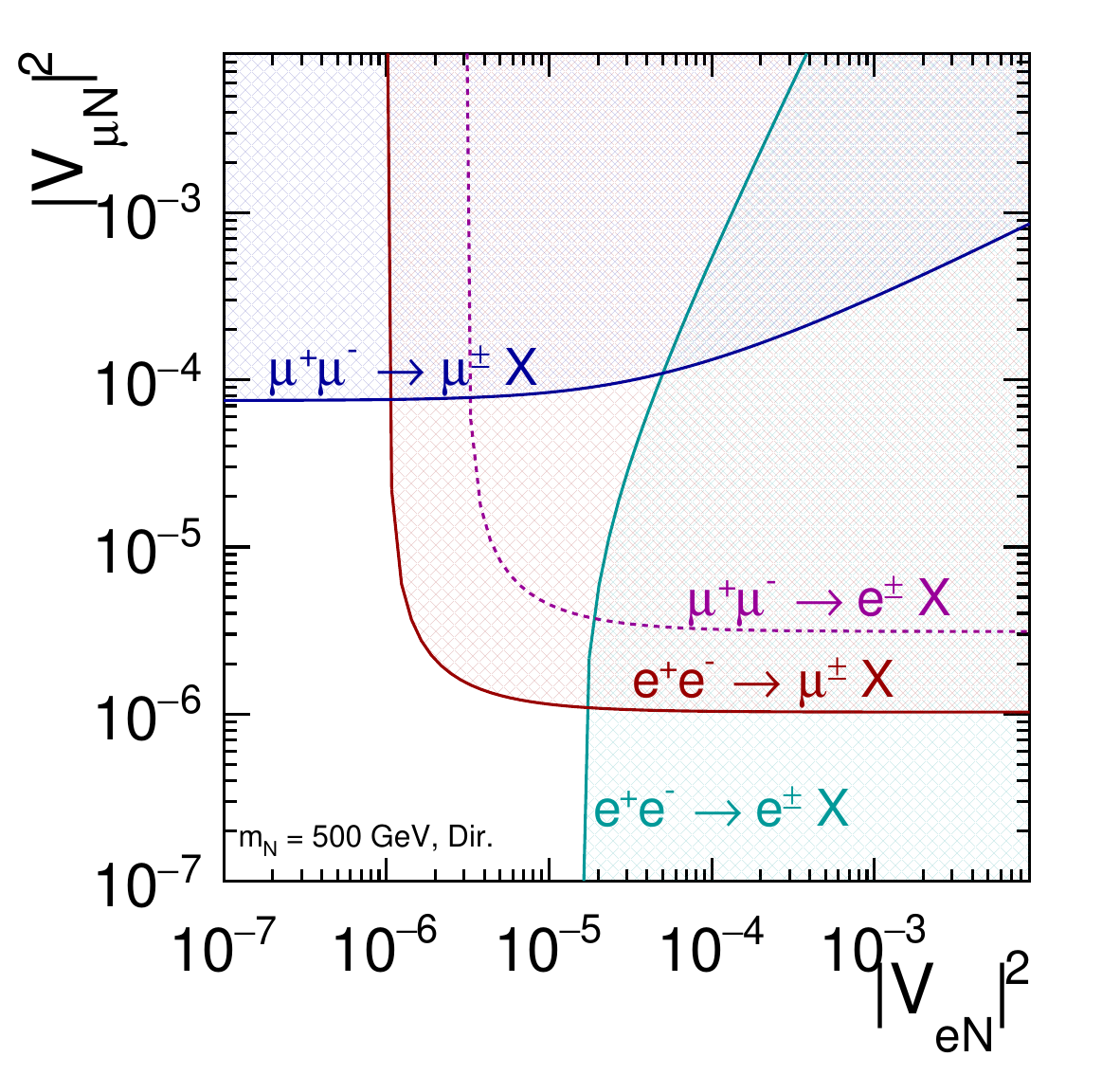}
    \caption{Dirac neutrino}
    \end{subfigure}
    \begin{subfigure}{0.49\textwidth}
    \includegraphics[width=\textwidth]{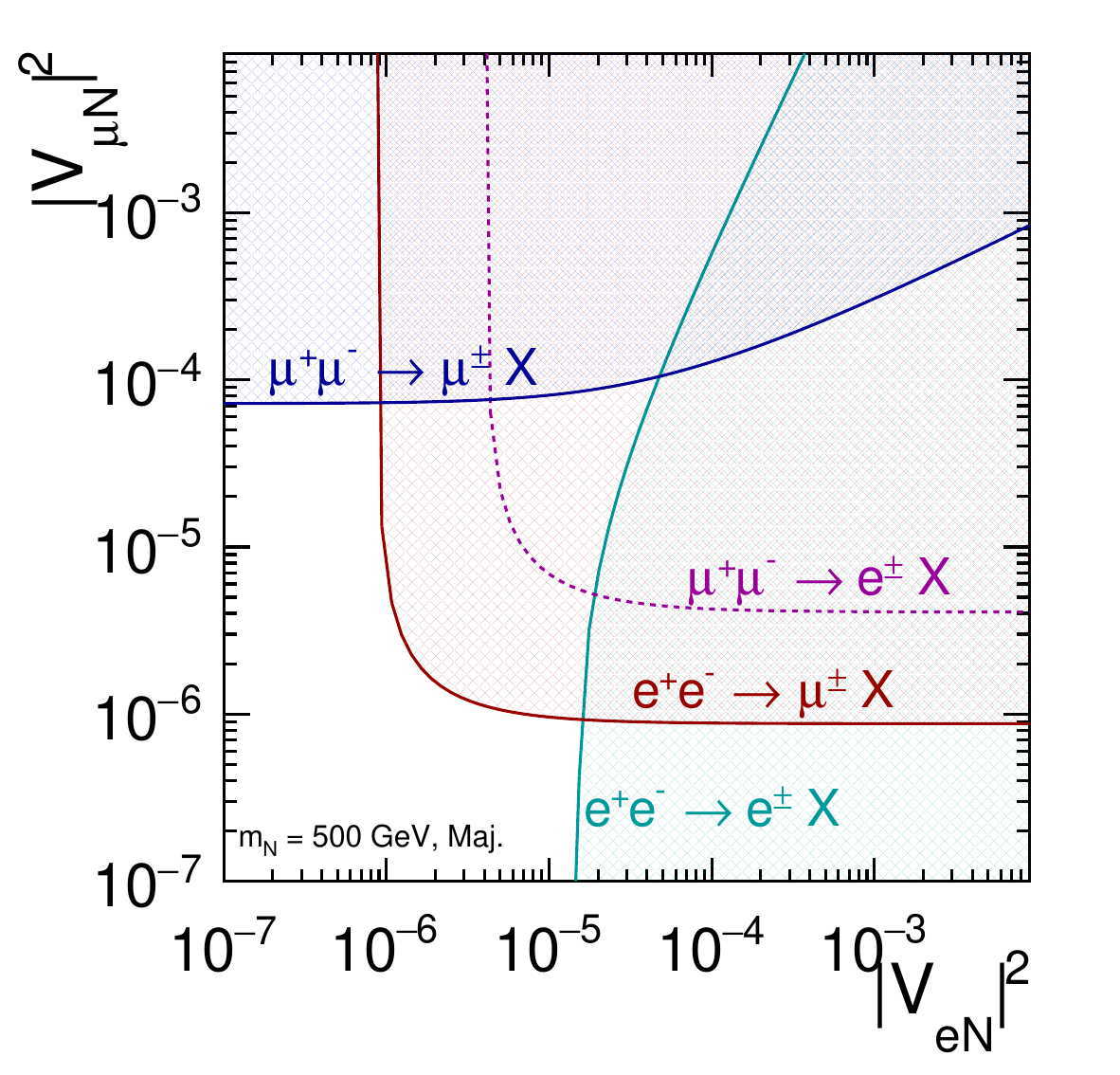}
    \caption{Majorana neutrino}
    \end{subfigure}
    \caption{Exclusion limits on the couplings $|V_{e N}|^2$ and $|V_{\mu N}|^2$ for a Dirac (left) and Majorana (right) neutrino with a mass of 500\,GeV expected for CLIC (denoted as $e^+e^-$) and Muon Collider ($\mu^+ \mu^-$), at 3\,TeV each, in the electron ($e^\pm X$) or muon ($\mu^\pm X$) channels. See text for details.}
    \label{fig:flavour}
\end{figure}

Given the expected exclusion reach, future lepton colliders appear to be an excellent place to study chiral properties of the heavy neutral leptons, which has been already mentioned in \cite{Mekala:2022cmm, Kwok:2023dck, Mikulenko:2023ezx}. However, a dedicated, experiment-oriented study on the discrimination of the heavy-neutrino nature has been missing in the literature. In Figure \ref{fig:flavour}, the exclusion limits on the couplings $|V_{e N}|^2$ and $|V_{\mu N}|^2$ based on the process $\ell^+\ell^- \to N \nu \to qq\ell \nu$ for Dirac and Majorana neutrinos with a mass of 500 GeV are compared. The plot was obtained for a simplified scenario, in which the tau coupling was set to zero\footnote{In fact, a similar plot could be obtained without this assumption, if a proper $\tau$-tagging procedure was employed. Even though the results are not expected to differ significantly, a dedicated study including full detector simulation would be required to account for $\tau$-tagging efficiency. In this paper, we refrained from elaborating on the issue in greater detail.}, by extracting separate results for CLIC and Muon Collider running at 3 TeV. Assuming that the process is mediated via t-channel $W$ exchange (which is a good approximation at this collision energy) and the heavy neutrino is a narrow resonance, the cross section scales as
\begin{equation}
    \sigma \sim \frac{|V_{\ell_b N}|^2 \cdot |V_{\ell_d N}|^2}{|V_{e N}|^2 + |V_{\mu N}|^2} \quad ,
\end{equation}
where $\ell_b$ and $\ell_d$ are beam and decay leptons, respectively. The numerical comparison using full flavor structure shows that the two different lepton machines are complementary in resolving the flavor structure of HNLs, while they cannot distinguish between Dirac and Majorana nature of the neutrinos in the current framework.  It clearly points to the necessity of establishing a dedicated discrimination procedure.

The paper is structured as follows:  we describe the model setup and simulation framework in Sections \ref{sec:model} and \ref{sec:setup}, respectively. In Section \ref{sec:analysis}, we present our analysis procedure and report on the results obtained. The most important features of the work are summarized in Section \ref{sec:conclusions}. We give extra technical details concerning the presented limit-setting procedure in Appendix \ref{sec:appendix}.

\section{Model setup}
\label{sec:model}
For the purpose of this study, we assume that heavy neutral leptons mixing with the SM partners are the only relevant trace of New Physics and no other new phenomena occur. Depending on the particular model, one may consider different numbers of non-standard neutrinos; however, for simplicity, we study the case where only one heavy neutrino with a mass ranging from 100 GeV to 10 TeV couples to SM particles. In fact, the assumption does not reduce the generality of the problem significantly, as it also covers the scenario in which more heavy neutral leptons are present but they are well-separated in mass (with respect to their width). Otherwise, the neutrino signal would be enhanced and its detection would be even less demanding.

Furthermore, all the heavy-neutrino couplings to the SM leptons (including $e$, $\mu$ and $\tau$) were set to the same value\footnote{As shown in Section \ref{sec:intro}, this assumption may be lifted when an interplay between $e^+e^-$ and $\mu^+\mu^-$ is considered.}. The Lagrangian for such a simplified model setup may be written as:
\begin{equation}
\Lagrangian[] 
= \Lagrangian[SM] + \Lagrangian[N] + \Lagrangian[WN\ell] + \Lagrangian[ZN\nu] + \Lagrangian[HN\nu]
\end{equation}
where $\Lagrangian[N]$ is a sum of kinetic and mass terms for heavy neutrinos (4-spinor notation is used everywhere):
\begin{equation}
\Lagrangian[N] = \xi_\nu \cdot  \left(  \bar{N}i\slashed{\partial}N - m_{N}\bar{N}N \right)
\end{equation}
with an overall factor $\xi_\nu = 1$ for the Dirac neutrino and $\xi_\nu = \frac{1}{2}$ for the Majorana neutrino scenarios.
$\Lagrangian[WN\ell]$ yields interactions with a $W$ boson:
\begin{equation}
\Lagrangian[WN\ell] = - \frac{g}{\sqrt{2}}W^{+}_{\mu} \sum^{\tau}_{l=e}\bar{N}V^{*}_{lN}\gamma^{\mu}P_{L}\ell^{-} + \textnormal{ h.c.},
\end{equation}
$\Lagrangian[ZN\nu]$ interactions with a $Z$ boson:
\begin{equation}
\Lagrangian[ZN\nu] = - \frac{g}{2\cos\theta_{W}}Z_{\mu} \sum^{\tau}_{l=e}\bar{N}V^{*}_{lN}\gamma^{\mu}P_{L}\nu_{l} + \textnormal{ h.c.},
\end{equation}
and $\Lagrangian[HN\nu]$ interactions with a Higgs boson, respectively:
\begin{equation}
\Lagrangian[HN\nu] = - \frac{gm_{N}}{2M_{W}}h \sum^{\tau}_{l=e}\bar{N}V^{*}_{lN}P_{L}\nu_{l} + \textnormal{ h.c.}
\end{equation}
The standard naming convention is used everywhere. Technical implementation was taken from the \textit{HeavyN} model~\cite{HeavyN}, developed for studies on Dirac~\cite{Pascoli:2018heg} and Majorana neutrinos~\cite{Alva:2014gxa, Degrande:2016aje}, using the UFO framework~\cite{Degrande:2011ua,Darme:2023jdn} .

\section{Simulation framework}
\label{sec:setup}
Following the approach of \cite{Mekala:2022cmm, Mekala:2023diu}, we studied processes of light-heavy neutrino associated production with the subsequent decay of the latter to two quarks and a lepton, $\ell^+ \ell^- \to N\nu \to qq\ell\nu$. This channel offers the possibility of the full reconstruction of the heavy neutrino from two jets and a lepton measured in the detector, if $\ell \in \left\{e, \mu\right\}$. As for the background, 4- and 6-fermion background processes with at least one lepton in the final state were considered. The simulation was performed at leading order in the SM couplings\footnote{Recently, NLO QCD and EW corrections have been automated~\cite{Bredt:2022dmm}, for the moment for the SM only, after earlier specially tailored attempts for single processes~\cite{Greiner:2011mp,Binoth:2009rv,Robens:2008sa,Kilian:2006cj}.} in \textsc{Whizard 2.8.5}~\cite{Moretti:2001zz,Kilian:2007gr} (ver. 3.0.0 was used for the Majorana signal generation). We generated reference samples with $V_{lN}^2 = 0.0003$ and all quark, electron and muon masses set to zero.
For the $e^+e^-$ colliders, the corresponding beam energy profiles were taken into account, as parametrized with the \textsc{Circe2} package, while the corresponding effect at the Muon Collider was neglected. There, one only has a small Gaussian beam energy spread which is irrelevant for this study. Parton shower and hadronization were modelled  with \textsc{Pythia~6}~\cite{Sjostrand:2006za}. 

Six different collider running scenarios were considered:
\begin{itemize}
\setlength\itemsep{-0.2em}
    \item ILC250 -- ILC running at 250 GeV, with an integrated luminosity of 900 fb$^{-1}$ and beam polarization of $-$80\% for electrons and +30\% for positrons,
    \item ILC500 -- ILC running at 500 GeV, with an integrated luminosity of 1.6 ab$^{-1}$ and beam polarization of $-$80\% for electrons and +30\% for positrons,
    \item ILC1TeV -- ILC running at 1 TeV, with an integrated luminosity of 3.2 ab$^{-1}$ and beam polarization of $-$80\% for electrons and +20\% for positrons,
    \item CLIC3TeV -- CLIC running at 3 TeV, with an integrated luminosity of 4 ab$^{-1}$ and beam polarization of $-$80\% for electrons (no polarization for positrons);
    \item MuC3TeV -- Muon Collider running at 3 TeV, with an integrated luminosity of 1 ab$^{-1}$ and no beam polarization,
    \item MuC10TeV -- Muon Collider running at 10 TeV, with an integrated luminosity of 10\,ab$^{-1}$ and no beam polarization.
\end{itemize}
The choice of the polarization setup was due to the chiral structure of the model, resulting in much higher cross sections for left-handed leptons and right-handed antileptons, but has little effect on signal-to-background discrimination.

Given the smallness of the electron mass, additional effects due to photon interactions need to be taken into account in the simulation chain for the $e^+e^-$ colliders. We considered both photon spectra from beamstrahlung and photons from collinear initial-state splittings (Weizsäcker-Williams Approximation~\cite{vonWeizsacker:1934nji, Williams:1935dka}) contributing to the following processes:
\begin{itemize}
\setlength\itemsep{-0.2em}
\item $\gamma e^{\pm}\to qq\ell$,
\item $\gamma\gamma\to qq\ell\nu$,
\item $\gamma\gamma\to qq\ell\ell$.
\end{itemize}
While beamstrahlung is absent at muon colliders, the corresponding initial-state splitting processes at the Muon Collider were found to have a negligible impact on the final results.

To account for detector effects, the framework for fast detector simulation \textsc{Delphes 3.5.0}~\cite{deFavereau:2013fsa} was employed in the next step. The default cards for each collider project were used for detector parameterization. Based on the expected signal topology consisting of one lepton and two reconstructed jets, we decided to use the exclusive two-jet clustering mode.

\section{Model-discrimination potential}
\label{sec:analysis}
As described in Section \ref{sec:intro}, search procedures are not expected to have any discriminative power between different possible natures of the heavy neutrino. Among the BDT variables used in \cite{Mekala:2022cmm, Mekala:2023diu}, the most important one was m$_{qql}$, the invariant mass of the dijet-lepton system, corresponding to the mass of the heavy neutrino. However, one may notice that this variable does not depend on the neutrino nature and thus, it does not hint towards any particular coupling structure. To address this issue, an extended set of analysis variables should be considered.

\begin{figure}[tb]
    \centering
    \includegraphics[width=0.49\textwidth]{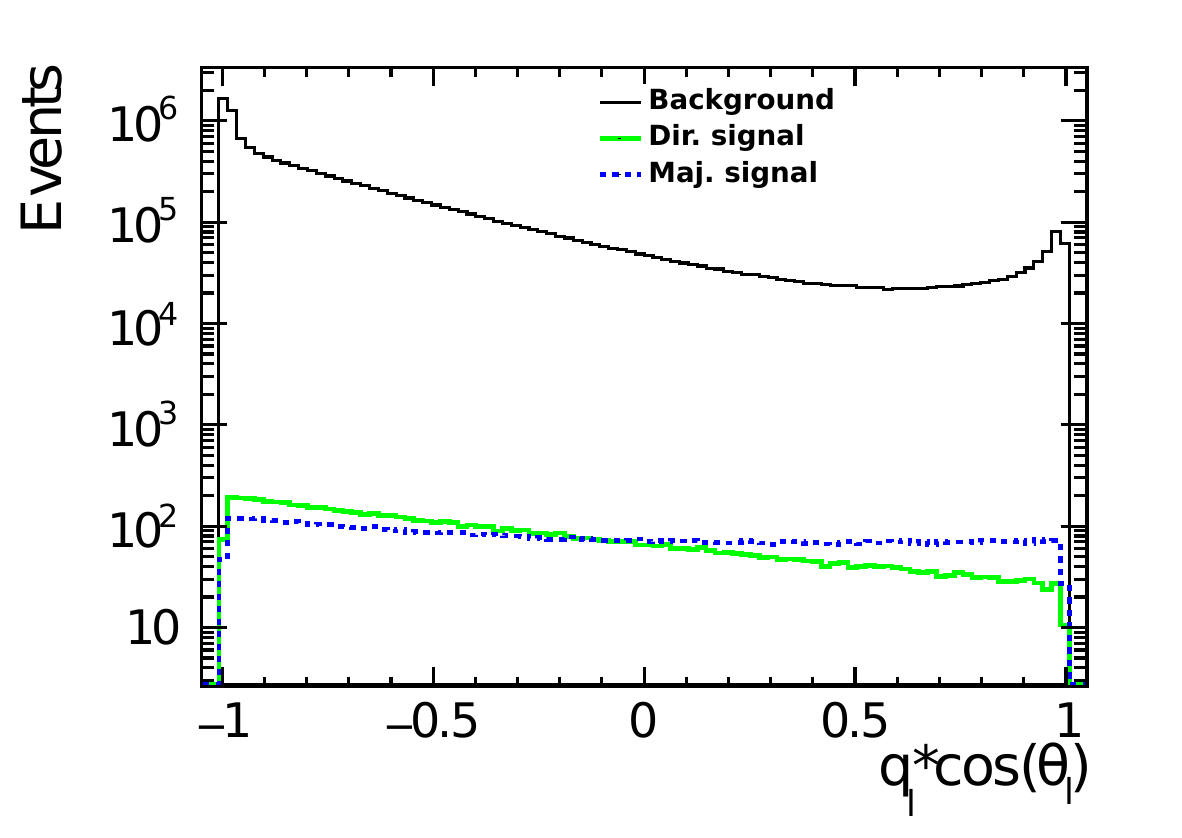}
    \includegraphics[width=0.49\textwidth]{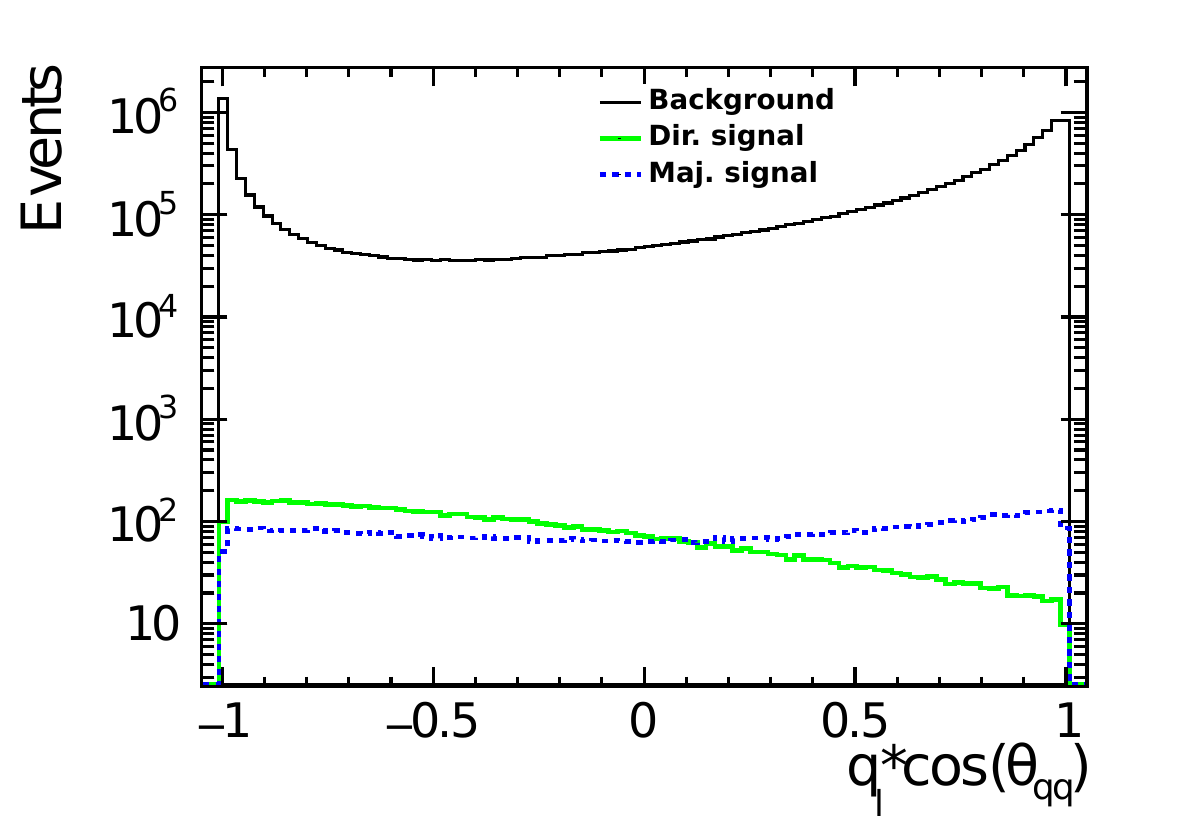}
    \caption{Distribution of the cosine of the lepton (left) and of the dijet (right) emission angle multiplied by the lepton charge for background (black solid line), Dirac signal (thick green line) and Majorana signal (dashed blue line) at ILC250. The heavy neutrino mass was set to 150 GeV.}
    \label{fig:variables}
\end{figure}

The difference between Dirac and Majorana particles lies in their CP properties. This means that for Majorana particles for any specific decay channel also its CP-conjugated one exists, leading to lepton-number violation, while for Dirac neutrinos only one of them (lepton-number conserving; this leads to the famous factor two in partial widths between the two cases). The chiral nature of weak decays together with the averaging over the decay process and its CP-conjugate for Majorana neutrinos leads to an experimental sensitivity, in particular, to the emission direction of a given final state particle (or anti-particle) in the rest frame of the decaying heavy neutrino. Hence, we decided to study kinematic variables incorporating this information. The 8-variable set developed for BDT training in \cite{Mekala:2022cmm, Mekala:2023diu}:
\begin{itemize}\setlength\itemsep{-0.2em}
\item m$_{qq\ell}$ -- invariant mass of the dijet-lepton system,
\item $\alpha$ -- angle between the dijet-system and the lepton,
\item $\alpha_{qq}$ -- angle between the two jets,
\item E$_{\ell}$ -- lepton energy,
\item E$_{qq\ell}$ -- energy of the dijet-lepton system,
\item p$^{T}_{\ell}$ -- lepton transverse momentum,
\item p$^{T}_{qq}$ -- dijet transverse momentum,
\item p$^{T}_{qq\ell}$ -- transverse momentum of the dijet-lepton system,
\end{itemize}
was extended by two new variables:
\begin{itemize}\setlength\itemsep{-0.2em}
\item q$_{\ell}\cdot \cos(\theta_\ell)$ --  cosine of the lepton emission angle multiplied by the lepton charge,
\item q$_{\ell}\cdot \cos(\theta_{qq})$ --  cosine of the dijet emission angle multiplied by the lepton charge.
\end{itemize}
The distributions of these variables, shown in Figure \ref{fig:variables}, behave differently for different event samples, offering a means of discrimination between the Dirac and Majorana natures of the heavy neutrino.

Using this extended set of 10 variables, the BDT training implemented in the \textsc{TMVA} package~\cite{Hocker:2007ht} was performed twice for each neutrino mass value, for the final-state lepton flavor experimentally favored at each collider type, i.e. the one for which the number of expected background events was lower: the muon channel for the $e^+e^-$ colliders and the electron channel for the Muon Collider, respectively. Based on the results of our previous studies \cite{Mekala:2022cmm,Mekala:2023diu}, we expect that contributions from the ``same-flavor'' channels would not change the results significantly.

\begin{figure}[t] 
         \centering
    \includegraphics[width=0.46\textwidth]{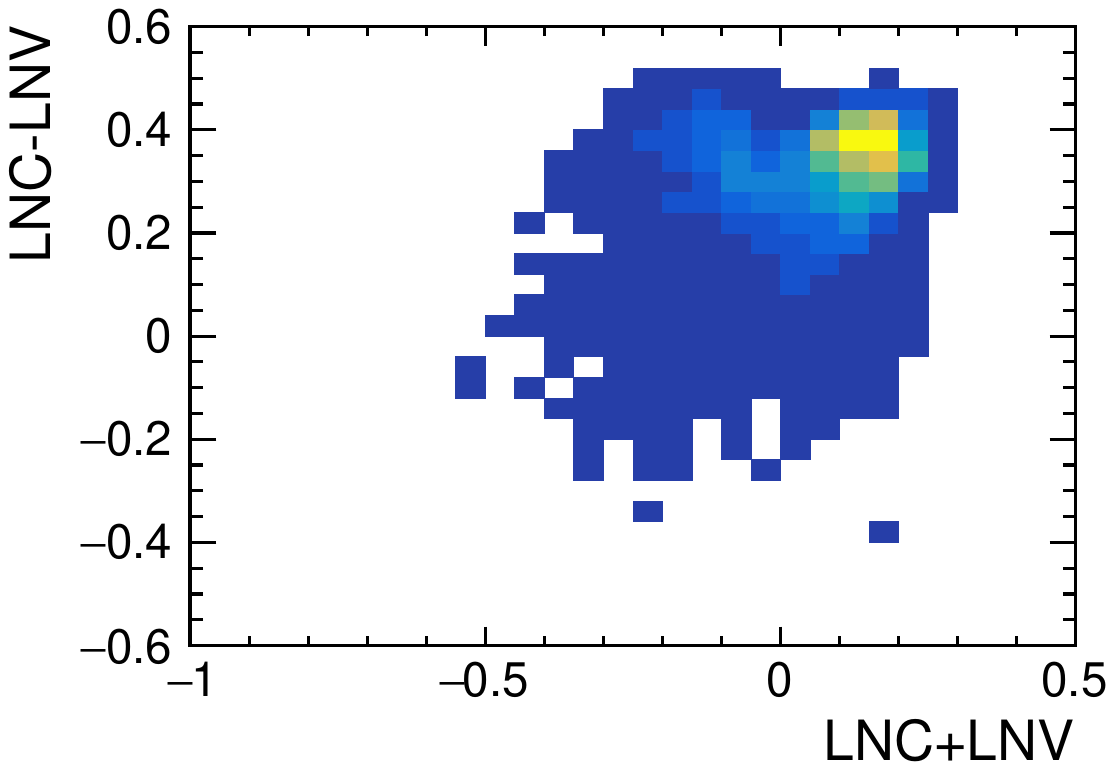} 
    \includegraphics[width=0.46\textwidth]{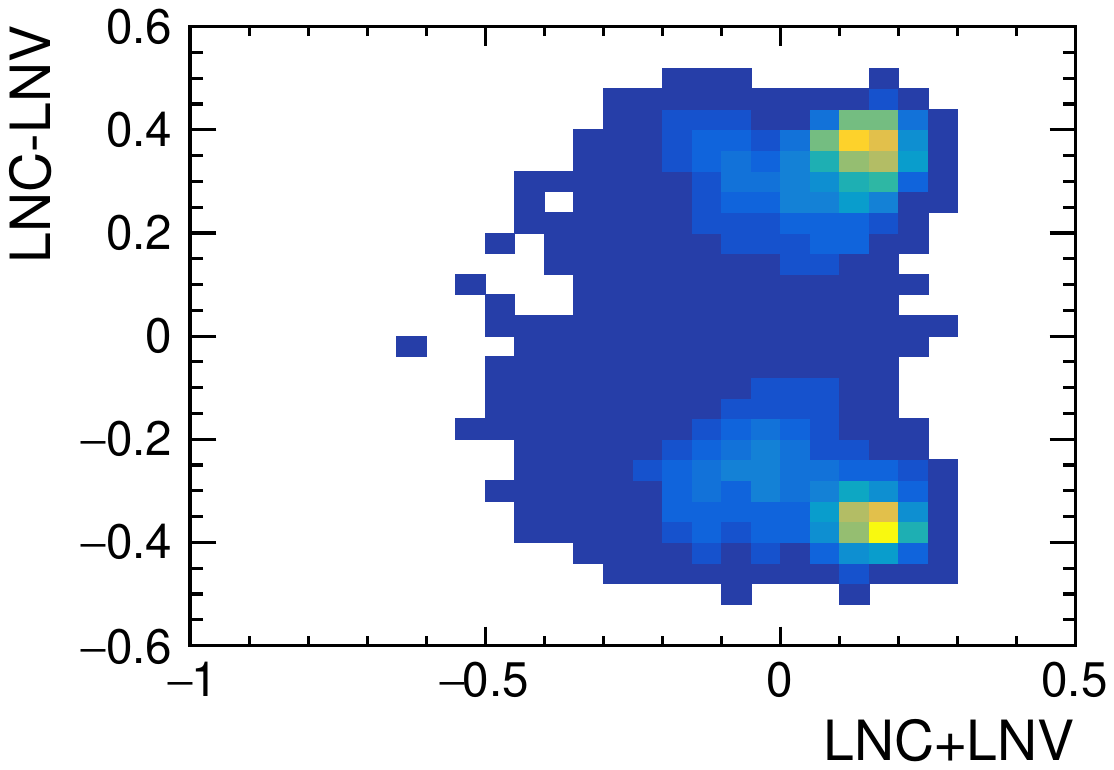}
    \includegraphics[width=0.46\textwidth]{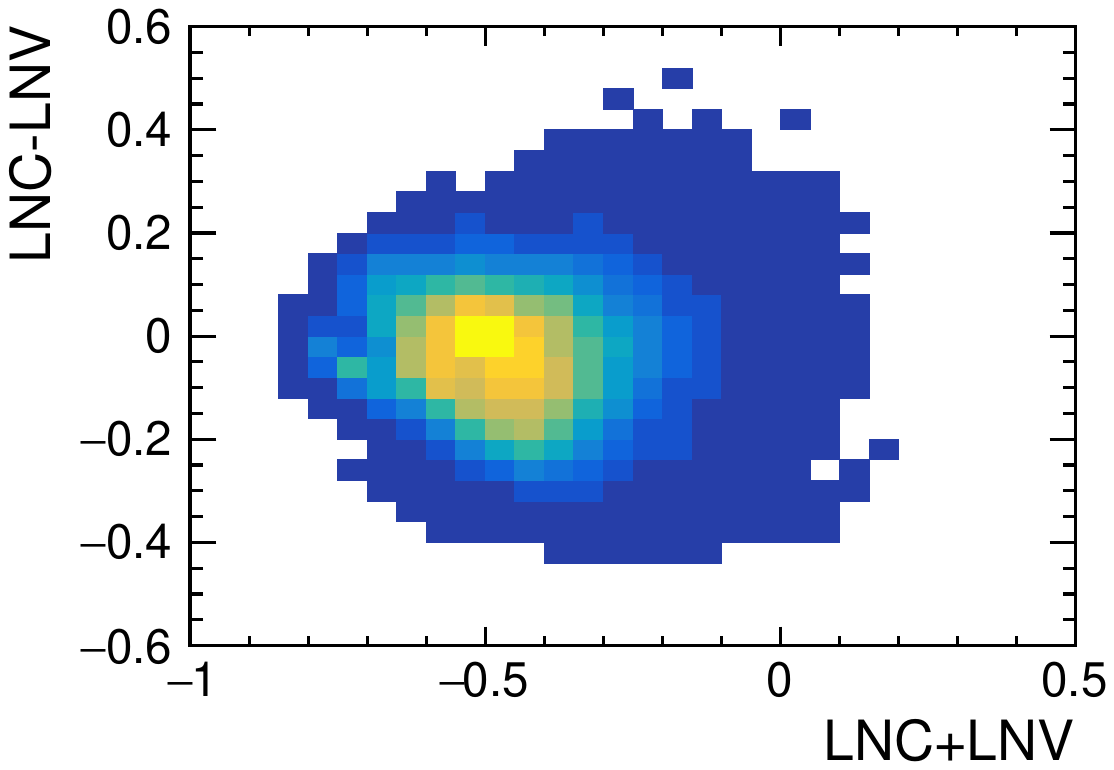}
    
    \caption{Two-dimensional distributions of the BDT classifier response for the Dirac neutrino signal sample (top left), the  Majorana neutrino signal sample (top right) and the background sample (bottom) for Muon Collider at 10 TeV. The mass of the heavy neutrino was set to 2 TeV and the weight parameter $\alpha_{BDT}$ to 1.}
         \label{fig:2dim}
\end{figure}

First, the algorithm was trained to distinguish between a signal sample of lepton-number-violating (LNV) heavy neutrino decays and a background sample contaminated with a lepton-number conserving (LNC) signal sample with some arbitrary weight $\alpha_{BDT}$. For instance, the parameter $\alpha_{BDT} = 1$ means that the relative weight of the SM background and the LNC signal sample is exactly the same as for the reference value of the mixing parameter. For the second training, the LNC decay sample was used as a signal and the LNV decay sample was used to contaminate the background. Then, 2-dimensional distributions of the sum and the difference of the two BDT responses, as shown in Figure \ref{fig:2dim}, were used for statistical analysis. The two-peak shape for the Majorana sample compared to the single-peak for the Dirac one clearly points to the specific chiral properties of the new particle.

In the next step, a $\chi^2$-like test statistic was introduced to compare the expected event distributions for different signal hypotheses:
\begin{equation}
    T' = \sum_{bins} \frac{[(B+D) -(B+M)]^2}{\frac{1}{2}[(B+D)+(B+M)]} = \sum_{bins} \frac{(D-M)^2}{B+\frac{D+M}{2}},
\end{equation}
where $D$, $M$ and $B$ are the numbers of Dirac, Majorana and background events in a given bin, respectively. The test statistic $T'$ was corrected to account for expected statistical fluctuations of the event numbers in the 2-D histogram bins:
\begin{equation}
    T' \to T = \sum_{bins} \frac{(D-M)^2}{B+\frac{D+M}{2}} + N_{\textnormal{dof}},
\end{equation}
where $N_{\textnormal{dof}}$ is the number of degrees of freedom (corresponding to the number of non-empty bins). To set final discrimination limits, we introduced a signal cross-section modifier, $\alpha_{lim}$, scaling the number of events in each bin for the signal samples:
\begin{equation}
    T(\alpha_{lim}) = \sum_{bins} \frac{\alpha_{lim}^2(D-M)^2}{B+\alpha_{lim} \cdot \frac{D+M}{2}} + N_{\textnormal{dof}}.
\end{equation}
To find the minimal coupling limit allowing for model discrimination at 95\% C.L., which we will refer to as the discrimination limit in the following, the parameter $\alpha_{lim}$ was varied for each mass to obtain the value of the test statistic $T$ corresponding to the critical value of the $\chi^2$ distribution for probability $p=0.95$ and the considered number of degrees of freedom: $T(\alpha_{lim}) = \chi^2_{crit}(N_{\textnormal{dof}})$. For the final discrimination, we decided to use $4 \times 4$ BDT bins which gave the most stable and also the most stringent limit. In order to optimize the BDT training, the procedure was iterated for different values of $\alpha_{BDT} \in \left\{0.1, 1, 10, 100\right\}$, producing a set of minimal values of $\alpha_{lim}$. To account for possible numerical fluctuations, each BDT training was repeated three times and the value of $\alpha_{lim}$ was averaged over runs. The smallest of the averaged values was used to scale the reference coupling parameter to obtain the final limit for each mass value. The validity of the technical procedure and our specific choice of setup are discussed in Appendix \ref{sec:appendix}.

\begin{figure}[t] 
    \centering
    \includegraphics[width=0.7\textwidth]{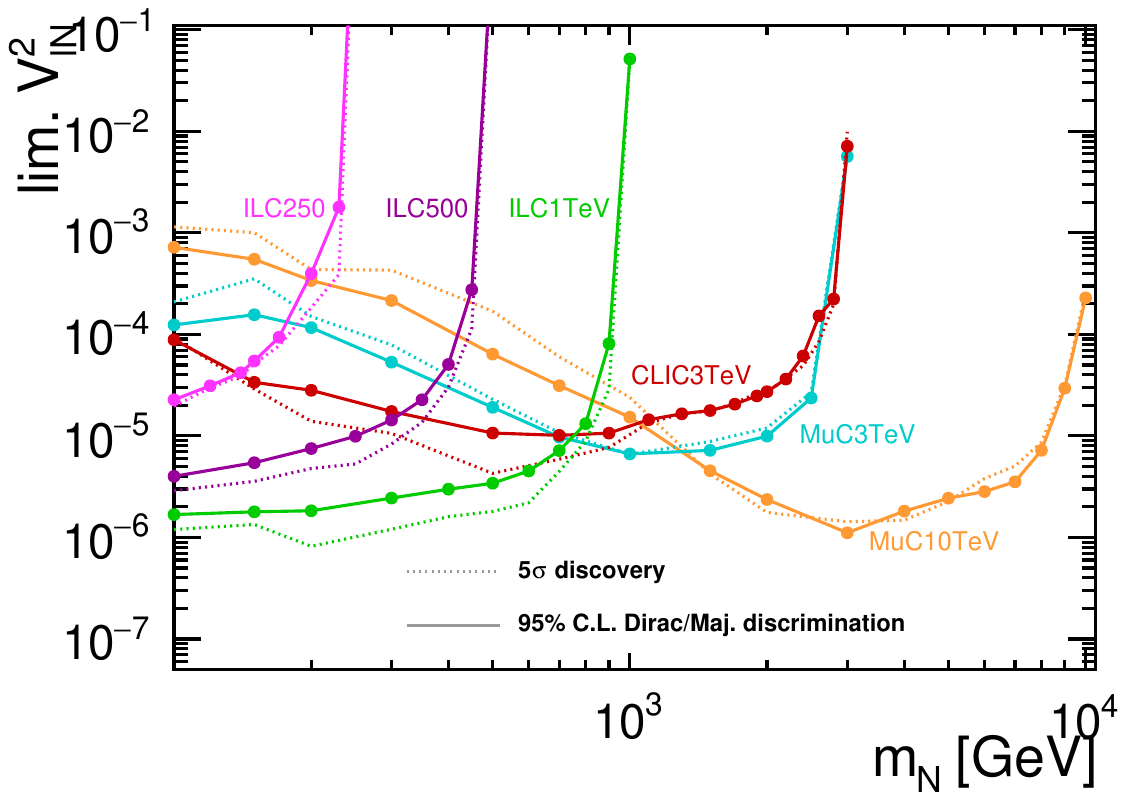}
    \caption{Comparison of the expected 95\% C.L. discrimination limits between Majorana and Dirac neutrinos (solid lines), as a function of the heavy neutrino mass, $m_N$, and the 5$\sigma$ discovery limits (dotted lines) for different collider scenarios considered in the study.}
    \label{fig:results_DvM}
\end{figure}

The final results of our study are shown in Figure~\ref{fig:results_DvM}. The 95\% C.L. discrimination limits are compared to the 5$\sigma$ discovery limits for the six collider scenarios considered in the study. The analysis confirms that once the heavy neutrinos are discovered at lepton colliders, it will be immediately possible to determine their nature (real or complex Lorentz representation) as well. 

\section{Conclusions}
\label{sec:conclusions}
Neutrinos are the most elusive particles of the Standard Model and many theories suggest that they can be a portal to New Physics. Heavy neutral leptons could potentially solve several problems of the fundamental theory and the proposed future lepton colliders (both electron-positron as well as muon colliders) seem to be best suited to search for their existence. We elaborated an analysis framework based on an extension of kinematic BDT variables to distinguish between a possible Dirac and Majorana nature of these particles. The crucial ingredient is variables combining kinematic information, like decay angles, with CP information, like the charge of the decay lepton. Our analysis shows that, by employing such variables encoding the chiral character of the particles, one may efficiently discriminate between complex and real Lorentz representations (i.e. Dirac or Majorana nature) of the heavy neutrino(s) at future lepton colliders simultaneously with their discovery.

\acknowledgments
The work was partially supported by the National Science Centre
(Poland) under the OPUS research project no. 2021/43/B/ST2/01778.
KM and JRR acknowledge the support by the Deutsche
Forschungsgemeinschaft (DFG, German Research Association) under
Germany's Excellence Strategy-EXC 2121 ``Quantum Universe''-3908333.  This work
has also been funded by the Deutsche Forschungsgemeinschaft (DFG,
German Research Foundation) -- 491245950.


\appendix
\section{Validation of the technical procedure of setting discrimination limits}
\label{sec:appendix}
In the following section, we 
give more details on the selected aspects of
the technical procedure of setting discrimination limits: we discuss the choice of the values of the training parameter $\alpha_{BDT}$ and of the number of bins used for BDT distributions.

\subsection*{Values of $\alpha_{BDT}$}
As explained in Section \ref{sec:analysis}, the parameter $\alpha_{BDT}$ was used to vary the relative weight of the background-contaminating ``wrong'' signal sample with respect to the ``proper'' background channels (not involving the heavy neutrino production) in order to optimize the BDT training for model discrimination. Nevertheless, the range of available values seems to be fully arbitrary. Intuitively, the values cannot be ``too small'' ($\alpha_{BDT} \to 0$) nor ``too large'' ($\alpha_{BDT} \to \infty$) -- in the former case, the classifier would be able to distinguish efficiently only between the signal sample and the ``proper'' background, while in the latter, only between the signal samples of the two different neutrino natures. The ideal classifier should, however, be able to effectively discriminate between three different classes: the Dirac signal, the Majorana signal and the background, and it can be optimized by a correct choice of $\alpha_{BDT}$. 

\begin{figure}[t] 
         \centering
    \includegraphics[width=0.6\textwidth]{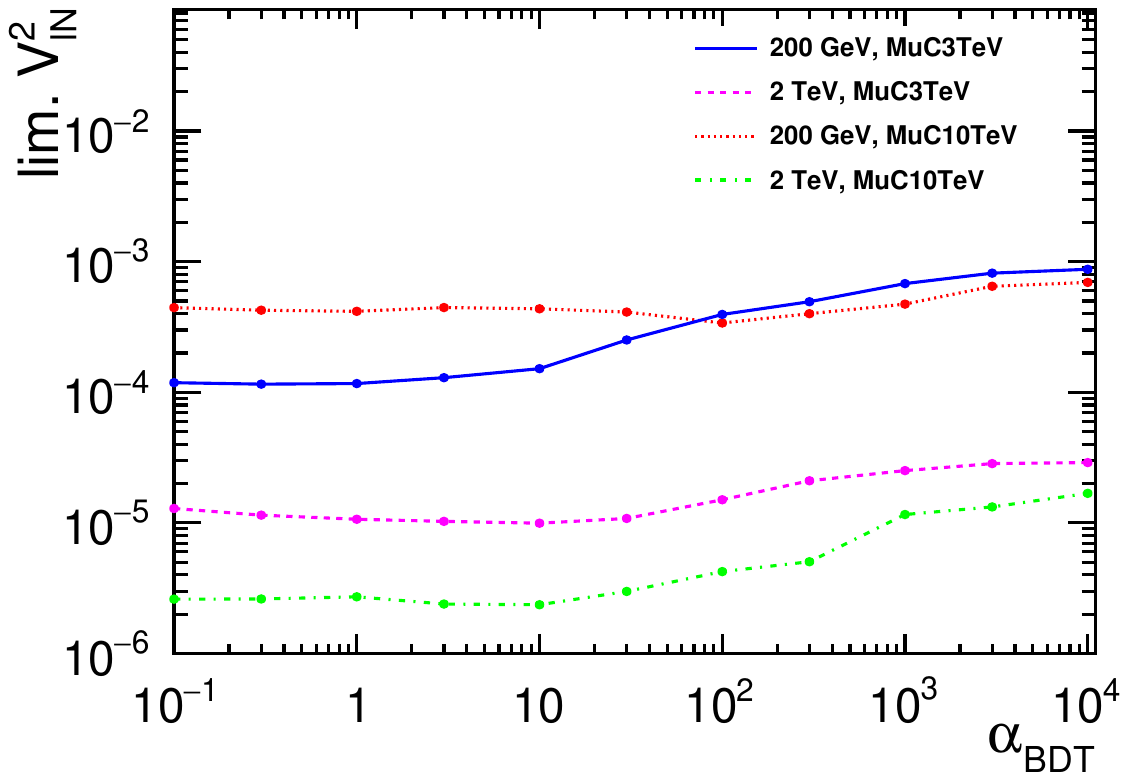}
    \caption{95\% C.L. discrimination limits for different values of the parameter $\alpha_{BDT}$ for two different masses of the heavy neutrino (m$_N = 200$ GeV and $2$ TeV) for two different running scenarios of the Muon Collider ($\sqrt{s}= 3$\,TeV and $10$\,TeV), as indicated in the plot.}
         \label{fig:alpha_BDT}
\end{figure}

In Figure \ref{fig:alpha_BDT}, we show 95\% C.L. discrimination limit results for four different experimental scenarios as a function of the value of the parameter $\alpha_{BDT}$. The results depend only mildly on the parameter -- for the vast range of the considered values, the results change by less than one order of magnitude. It proves that the proposed procedure is stable and gives reliable limits even if the choice of $\alpha_{BDT}$ may seem to be formally arbitrary.

\subsection*{Number of bins}
\begin{figure}[t] 
         \centering
    \includegraphics[width=0.6\textwidth]{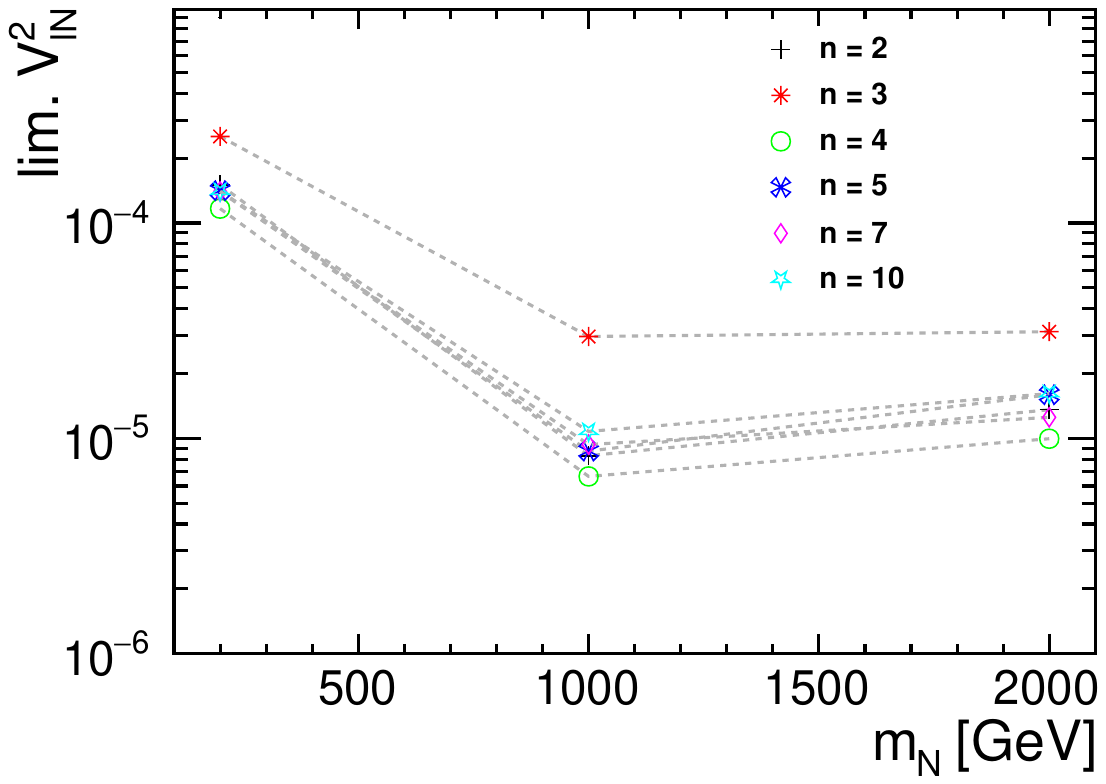}
    \caption{95\% C.L. discrimination limits for different masses of the heavy neutrino at the Muon Collider running at 3 TeV; different point styles stand for different numbers of bins ($n \times n$) used for the $\chi^2$ statistic.}
         \label{fig:bins}
\end{figure}

Similarly, the choice of the number of bins entering the $\chi^2$ test statistic is nontrivial. Naively, a larger number would ensure more information about the sample classification. Nonetheless, by increasing the number of bins, one would also increase the number of degrees of freedom and hence, the critical value of the $\chi^2$ distribution and deplete the event number per bin. Thus, probing different setups was crucial to optimize the analysis. A simplified validation test is shown in Figure \ref{fig:bins} where discrimination limits for three different masses were obtained using different numbers of bins. It is visible that the $4 \times 4$ case slightly supersedes other choices. One may also notice that only the $3 \times 3$ binning performs significantly worse than other options. The issue may be understood by studying the distribution in Figure \ref{fig:2dim}. Even though the three distributions are distinguishable, the maxima are not distant from the central value. By employing the $3 \times 3$ binning (9 bins in total), the significant deviation of the distributions is hidden in the central bin, resulting in the less stringent limit. It was verified that similar behavior holds for other masses and collider scenarios.

\bibliographystyle{JHEP}
\bibliography{biblio.bib}

\end{document}